\documentclass[aps,amsmath,amssymb,reprint]{revtex4-2}

\usepackage{graphicx}
\graphicspath{ {./figures/} }
\usepackage{tikz}
\usepackage{amsmath}
\usepackage{dcolumn}
\usepackage{bm}
\usepackage{color,hyperref}
\usepackage[english]{babel}
\usepackage{natbib}
\usepackage{float}
\usepackage{media9}
\usepackage{caption} 
	\definecolor{LinkColor}{rgb}{0.5,0.5,0}
	\definecolor{UrlColor}{rgb}{0,0.5,0.7}
	\definecolor{CiteColor}{rgb}{0.5,0.5,0.5}
	\definecolor{olivegreen}{RGB}{0, 70, 0}
    \definecolor{mark}{RGB}{205,222,255}
	\hypersetup{%
	linkcolor=LinkColor,%
	filecolor=LinkColor,%
	menucolor=LinkColor,%
	citecolor=CiteColor,%
	urlcolor=UrlColor,%
	colorlinks=true%
	}

\newcommand{\hc}{$\mathrm{H_{c}}$}
\newcommand{\Tc}{$\mathrm{T_{C}}$}
\newcommand{\dg}{$^{\circ}$}
\begin{document}
	
\title{Magnetic domains in ultrathin, bulk-like and proximity-coupled Europium Oxide}
\author{Seema}\email{seema.seema@uni-konstanz.de}
\author{Moumita Kundu}
\author{Paul Rosenberger}
\author{Henrik Jentgens}
\author{Ulrich Nowak}
\author{Martina M{\"u}ller}\email{martina.mueller@uni-konstanz.de}
\affiliation{Fachbereich Physik, Universit{\"a}t Konstanz, 78457 Konstanz, Germany}

\date{\today}
\begin{abstract}

The control of electron spins in materials that are simultaneously ferromagnetic and insulating opens up a wealth of quantum phenomena in spin-based electronics. Thin films of europium oxide (EuO) are ideal for the generation and manipulation of  spin-polarized states, but so far there are no experimental literature reports on the magnetic domain patterns for EuO. However, at these microscopic length scales, magnetic relaxation between the remanent and demagnetized states takes place in any spintronic device. This relaxation process involves displacements of magnetic domain walls and can therefore be strongly influenced by the film structure and thickness. 

Here we present an investigation of the temperature-dependent behavior of magnetic domains and hysteresis in bulk-like (25 nm) and ultrathin (3 nm) EuO films. Magneto-optical Kerr microscopy is used, a technique that is a valuable tool to explore microscopic features such as spin dynamics and magnetic domain walls. 
Significant Kerr rotation in EuO led to high-contrast magnetic domain images in thick films, facilitating observation of domain dynamics. The critical temperature (\Tc) and coercivity shows strong thickness-dependent variations. The analysis and comparison of hysteresis loops and domain imaging in EuO and EuO/Co reveal proximity effect-induced antiferromagnetic coupling of both layers. To elucidate the magnetization reversal dynamics in EuO, micromagnetic simulations using MuMax$^3$ were performed below and above \Tc. This comprehensive approach aims to comprehend the impact of magnetism and magnetic proximity effect in EuO on the micromagnetic scale, potentially extending its magnetic ordering beyond \Tc.

\end{abstract}
	
\maketitle

\section{\label{sec:intro}Introduction}
	
Magnetic insulators are ubiquitous materials in spintronic devices\,\cite{Taroni2013, EuO_book, Parkin2008, Caspers2016, Mueller2011}. The material class of rare-earth europium chalcogenides (EuX, X\,$=$\,O, S, Se, and Te) features strong optical responses and enormous magneto-optic Kerr effect that opens up spin-based applications including nanophotonic devices, e.g. spin diodes, spin transistors, spin injectors, spin qubits and non-volatile magnetic memory\,\cite{Kats2023,pappasEuSCo}. Among them, EuO has attracted the most attention in the past couple of decades due to its Silicon-like bandgap of 1.12\,eV and a reasonable Curie temperature (\Tc)\,of 69\,K. Below \Tc, the magnetic moments of Eu$^{2+}$ are exchange-coupled with a very large localized magnetic moment of 7.0\,$\mu_\text{B}$. Because of the significant exchange splitting of the conduction band, EuO also shows a strong spin polarization near 100\,$\%$. EuO has the potential to be topologically driven materials with significant spin-orbit coupling, as evidenced by several observed phenomena such as anomalous Hall effect and presence of magnetic skyrmions\,\cite{Ohuchi2015}.
	
In the wealth of literature available on EuO, most of the studies on magnetic properties have been dedicated to its macroscopic characteristics such as \Tc, electronic structure and temperature-dependent magnetization~\cite{Heider2022, Prinz2016}. The practical application of EuO is restricted due to its \Tc~of 69\,K and experimental efforts are dedicated to enhance the \Tc. The improved exchange coupling by the donor states and/or conduction electrons with electron doping via oxygen vacancies or trivalent rare-earth (RE) ions\,\cite{JMMM_EB_EuO, offstoiEuO, EuOGaN_NM2007} were the sources of the rise in \Tc. Doping EuO with RE such as Lu, Gd, and La had resulted in \Tc\,values of 119, 116, and 120\,K\,\cite{LaEuO2010, LuEuO2012, GdEuO}. Proximity effects by placing a room temperature ferromagnet close to EuS has been investigated by Pappas et al.\,\cite{pappasEuSCo} and because of Co, thin EuS layers showed significantly high spin polarization up to room temperature (RT). Using the proximity-induced effect from Co and Fe, this strategy has been tested for ultrathin EuO (3 nm); very lately, signatures of magnetic ordering till 80\,K have been observed~\cite{HAXPES_MCD_EuO}. However, all of these reports are confined to the physical properties in the bulk or nano regime. 

Given this, the microscopic and mesoscopic properties such as magnetic domains, magnetization reversal, and domain dynamics in EuO are unknown yet. There is a dearth of literature on magnetic domains in EuO and one reason for this could be the oxidation-prone nature of EuO to the more stable paramagnetic phase Eu$_2$O$_3$\,\cite{aewski2021}. Despite the fact that EuO has the largest Kerr rotation of $\sim$\,4\,degrees, there are a only handful of reports on Kerr studies of EuO\,\cite{greiner1966}. Some of these focus on the energy-dependent Kerr rotation of Eu\,\cite{greiner1966, Brullion_FM_Heisenberg, MM_JAP_2009, Kats2023}, while the others are related to the second harmonic generation in EuO\,\cite{Matsubara2010}. Recently, M{\"o}nkeb{\"u}scher et. al. studied EuO/Co bilayers using static and time-resolved magneto-optical Kerr effect measurements, and explored the influence of magnetic proximity of Co on the \Tc\,and on the spin dynamics in EuO\,\cite{dortmund2023}. 

The present work focuses on the coercivity behavior and magnetic domain reversal studies of high-quality stoichiometric EuO films with different thicknesses from the bulk-like (25\,nm) to the ultrathin films regime (3\,nm). Using magneto-optic Kerr microscopy, we simultaneously observed magnetic domains and recorded magnetic hystereses on EuO films. We also investigated the effects of temperature and film thickness on the coercivity of EuO. Moreover, magneto-optic studies have been conducted on a EuO/Co heterostructure to examine the impact of the close proximity of Co on the magnetic domains and saturation behavior of EuO. To picture the magnetization reversal dynamics in EuO and EuO/Co, micromagnetic simulations were performed using MuMax$^3$ and compared with the experimental findings.

\section{\label{sec:exp}Experimental}
\subsection{\label{sec:refexp}Methods}

Using molecular beam epitaxy (MBE), EuO thin films and EuO/Co heterostructures were deposited on Nb-doped SrTiO$_3$ (Crystec GmbH). A 25\,nm thick EuO film was deposited as a bulk reference at 400\dg\,C at a Eu deposition rate maintained at 0.05\,Å/s  and with an oxygen partial pressure of 10$^{°-7}$\,mbar after 30 minutes of substrate preheating at 500\,\dg\,C. For thinner 3\,nm EuO film a redox growth mechanism was used. More information about the redox growth mechanism can be found elsewhere~\cite{PL_MM_PRM2019, PR_MM_PRM_2022}. Then, using e-beam evaporation, a 4\,nm thick Co overlayer was deposited onto the 3\,nm EuO film at room temperature. Co was deposited on half of the sample area using a shadow mask. For preliminary structural and chemical examination, all samples were subjected to low energy electron diffraction and in situ X-ray photoelectron spectroscopy (XPS) (not covered in this study). Lastly, a 25\,nm coating of MgO was applied to all samples to prevent air oxidation.\\
A wide-field Kerr microscope from Evico Magnetics, Germany, was used for the magnetic measurements. It employs a microscope of the Carl Zeiss Axioscope design, equipped with a brilliant white LED light source. The longitudinal geometry measurements (longitudinal magneto-optic Kerr effect, or LMOKE) were made using an applied field that was variable between $\pm$300\,mT. By fully magnetizing the sample to saturation, i.e. ideally to a single domain state, a background image with domain information is subtracted from it to get pure domain contrast, free of topographic influences. The background picture is subtracted from the live image during each magnetic field step to obtain a domain image at the applied field.

\subsection{\label{sec:refexp}Simulations}

Micromagnetic simulations were performed using Mumax$^3$ to obtain quantitative information about the reversal mechanisms in EuO~\cite{Vansteenkiste2014, DeClercq2017}. We assumed a micromagnetic model to reproduce the magnetization properties of an experimentally observed thin EuO film. For the sake of simplicity, the crystalline texture of the thin film was neglected. The field-dependent curves for EuO were obtained along with snapshots at each value of the applied field. The magnetic parameters were kept close to the theoretical values for EuO to correlate to the respective experimental results by MOKE. The following set of parameters was used in the simulation process: M$_S$ = 1.77{$\times$}10$^6$\,J/T, K$_C$=-43600\,J/m$^3$, and A$_{exch}$=1.1658{$\times$}10$^{-12}$\,J/m, where M$_S$, K$_C$ and A$_{exch}$ are the saturation magnetization, cubic anisotropy constant, and exchange constant, respectively~\cite{Trepka2017, miyata1967}.
	
\section{\label{sec:results}Results}
	
\subsection{\label{sec:refEuO}Bulk-like EuO film}
	
\begin{figure*} 
\includegraphics[width=145mm]{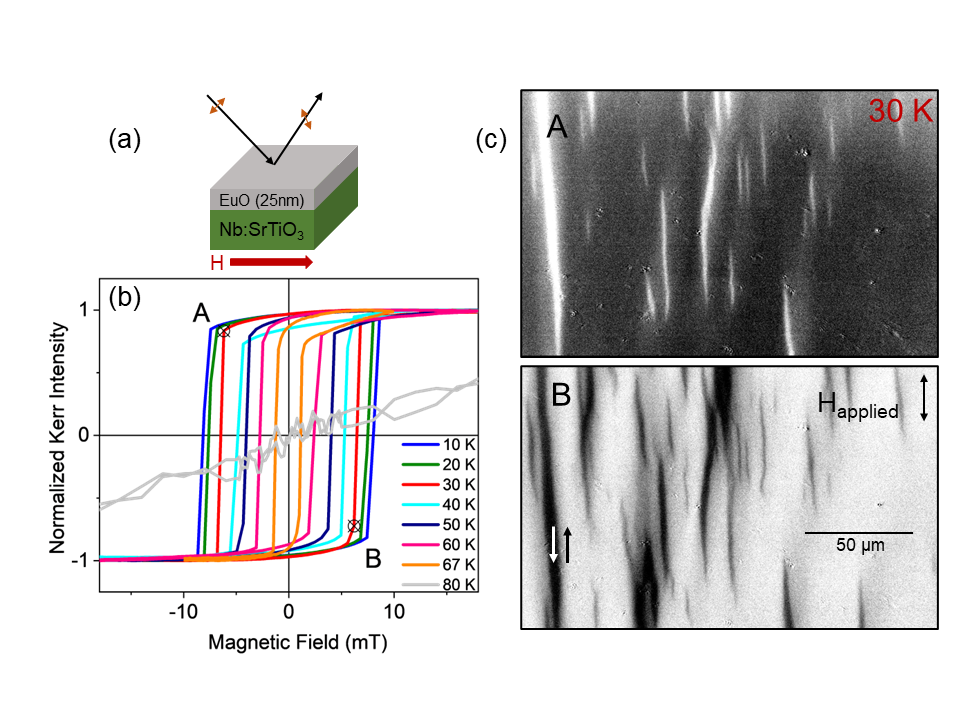}
\vspace{-10mm}
\caption{\label{fig:fig1}(a) Schematic showing the sample structure and LMOKE geometry, (b) Normalized magnetic hystereses obtained on the 25\,nm EuO reference film using LMOKE at different temperatures, (c) shows 180\,\dg\,stripe magnetic domains observed at different acquisition points A and B of the hysteresis measured at 30\,K shown in (b).}
\end{figure*}

First, we present the magnetic hystereses measured for the 25\,nm EuO bulk-like film. We recorded magnetic domain pictures in LMOKE geometry (Fig.~{\ref{fig:fig1}}~(a)) and measured magnetic hysteresis by systematically varying the sample temperature below and up to \Tc. The LMOKE hystereses are displayed in Fig.~{\ref{fig:fig1}}~(b). Well-defined, fully saturated hystereses were observed at every measured temperature up to \Tc. The coercivity (\hc) of the loops decreases with increasing temperature, as is expected for ferromagnetic behavior\,\cite{Chen_SW_model}. The value of \hc\,= 7.5\,mT at 20\,K is in good agreement with the values reported in the literature, which are characteristic of high-quality EuO films measured with vibrating sample magnetometry and MOKE\,\cite{dortmund2023}. As the temperature approaches the bulk EuO \Tc~i.e. 69\,K, both the \hc\,and remanent magnetization fall drastically. The magnetic hysteresis could be observed up to 70\,K, which corresponds to the bulk \Tc~of EuO.

Fig.~{\ref{fig:fig1}}~(c) shows the representative domain images observed at point A and B during the measurement of LMOKE hysteresis at 30\,K as indicated in Fig.~{\ref{fig:fig1}}~(b). In point A, the light gray contrast show the emerging areas as stripes which are aligned in the direction of the applied field. These are 180\,\dg\,domains and they start to appear as stripes or branches in the direction of the applied field at different regions in the measured area. This could be attributed to the presence of defects in the film that act as pinning centers for the domain nucleation. The width of these stripes is $\approx$\,5\,$\mu$m and increases as domains grow in size due to the strength of the applied field. A complete reversal of the described behavior can be seen when the applied field is varied from + to - direction i.e. at point B in Fig.~{\ref{fig:fig1}}~(c). The process of magnetic reversal and evolution of magnetic domains can be seen in the multimedia file in Fig.~2S in the supplementary information. A similar kind of domain pattern was observed at all measurement temperatures, while the Kerr contrast gradually reduced until the signal was lost when \Tc\,was approached, as shown in Fig.~{\ref{fig:fig1}}~(b).

\subsection{\label{sec:EuOCo}Ultrathin EuO and EuO/Co heterostructure}
	
\begin{figure*}
\centering
\includegraphics [width=145mm] {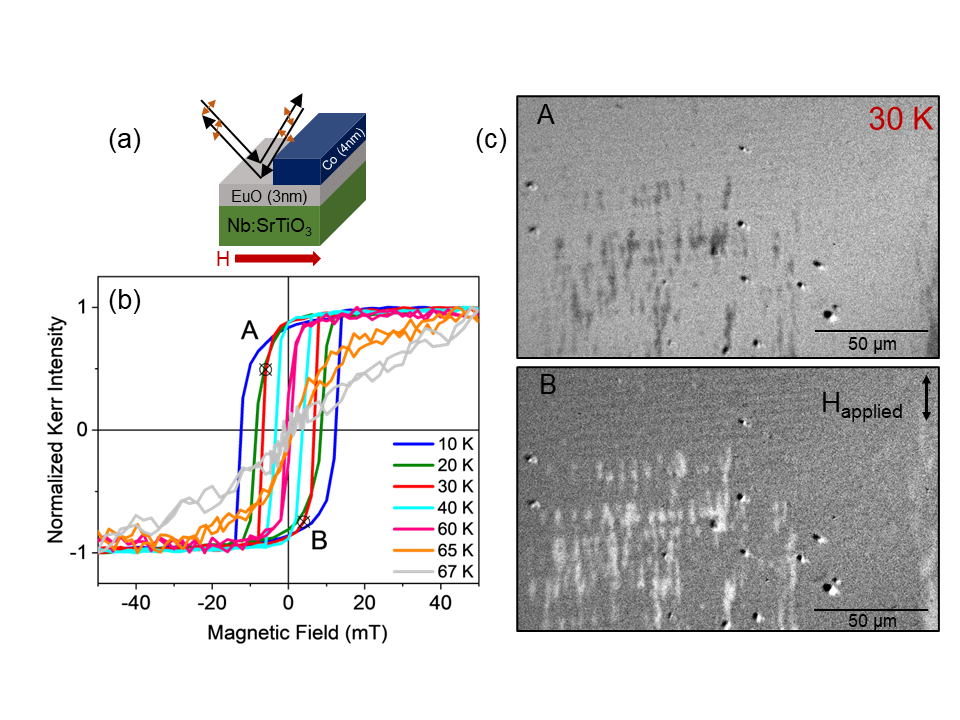}
\vspace{-10mm}
\caption{\label{fig:fig2} (a) Schematic showing the two-sided EuO and EuO/Co heterostructure measured in pure LMOKE geometry. (b) Normalized magnetic hystereses obtained on a 3\,nm EuO film using pure LMOKE at different temperatures. The magnetic domain observed at different acquisition points (A and B) of a hysteresis curve measured at 30\,K for a 3\,nm EuO film. The light and dark gray color represents the regions where the magnetization is oppositely aligned.}
\end{figure*}

We performed similar LMOKE measurements on both the EuO and Co sides of the EuO/Co heterostructure as shown in schematic Fig.~{\ref{fig:fig2}}~(a). However, since the EuO film is only 3\,nm in thickness and the corresponding Kerr signal obtained was much lower than for the 25\,nm bulk reference, we performed the experiment as a pure LMOKE hysteresis measurement to enhance the signal-to-noise ratio as described in the supplementary information. On the EuO ultrathin film side, the hystereses obtained as a function of temperature are shown in Fig.~{\ref{fig:fig2}}~(b). The hystereses showed lesser magnetic remanence (0.83) as compared to the reference film (0.95) at 10\,K, as can be seen in Fig.~1S~in the supplementary information. At 10\,K, \hc\,is 12.5\,mT, which is higher than the value for the reference 25\,nm film. The remanent magnetization remains almost similar to the saturation as the sample temperature is increased, the \hc\,decreases, showing similar behavior as observed for the reference film. A comparison of the shape of hysteresis obtained at 10\,K for both EuO films can be seen in the supplementary information. The 3\,nm EuO film starts to lose its Kerr signal at lower temperature than 25\,nm EuO and exhibits a very weak Kerr signal at and above 65 K, which is due to reduced magnetic exchange interaction in EuO in the atomic layer limit. The behavior is consistent with the modeling of thickness-dependence of \Tc\,in EuO films as shown by M{\"u}ller et al.\,\cite{MM_JAP_2009} and in other reports as well\,\cite{Brullion_FM_Heisenberg, thicknessEuO}: The \Tc\,decreases with decreasing film thickness, and comparatively lower \Tc\,values have been observed for the EuO films thinner than 5\,nm. This is in accordance with the hystereses measurement for the 25\,nm and 3\,nm EuO film.

The magnetic domain images corresponding to point A and B on the loop measured at 30\,K for a 3\,nm EuO sample are shown in the Fig.~{\ref{fig:fig2}}~(c). Similar to Fig.~{\ref{fig:fig1}}~(c), the domain evolution for the 3\,nm EuO film can be observed in picture A of Fig.~~{\ref{fig:fig2}}~(c). Here the magnetic domains clearly show a weaker Kerr contrast as compared to Fig.~\ref{fig:fig1}~(c). Also, the stripes observed were smaller in size ($\approx$\,2-5\,$\mu$m) and domains have simultaneously more nucleation sites. A similar domain pattern with reversed Kerr contrast could be obtained at point B. Comparing the domain images of two different thicknesses of EuO, 3 and 25\,nm, we could observe subtle differences in particular such as: (i) the domains in the 3\,nm EuO film were smaller in size, (ii) possess less Kerr intensity contrast, (iii) had more number of nucleation sites. These observations can be understood by comparison of thickness-dependent thin film properties such as crystallite size, microstrain, concentration of defects, etc.\,\cite{Miao2005}. These structural properties are strongly dependent on the thickness of the film and can alter the magnetic anisotropy and the domain evolution pattern as well\,\cite{Miao2005, JENSEN2006, Poulopoulos1999}.  A short video clip of the process of domain evolution and magnetic reversal can be seen in the multimedia file in the Fig.~3S of the supplementary information.
	
\begin{figure} \centering
\includegraphics [width=95mm]{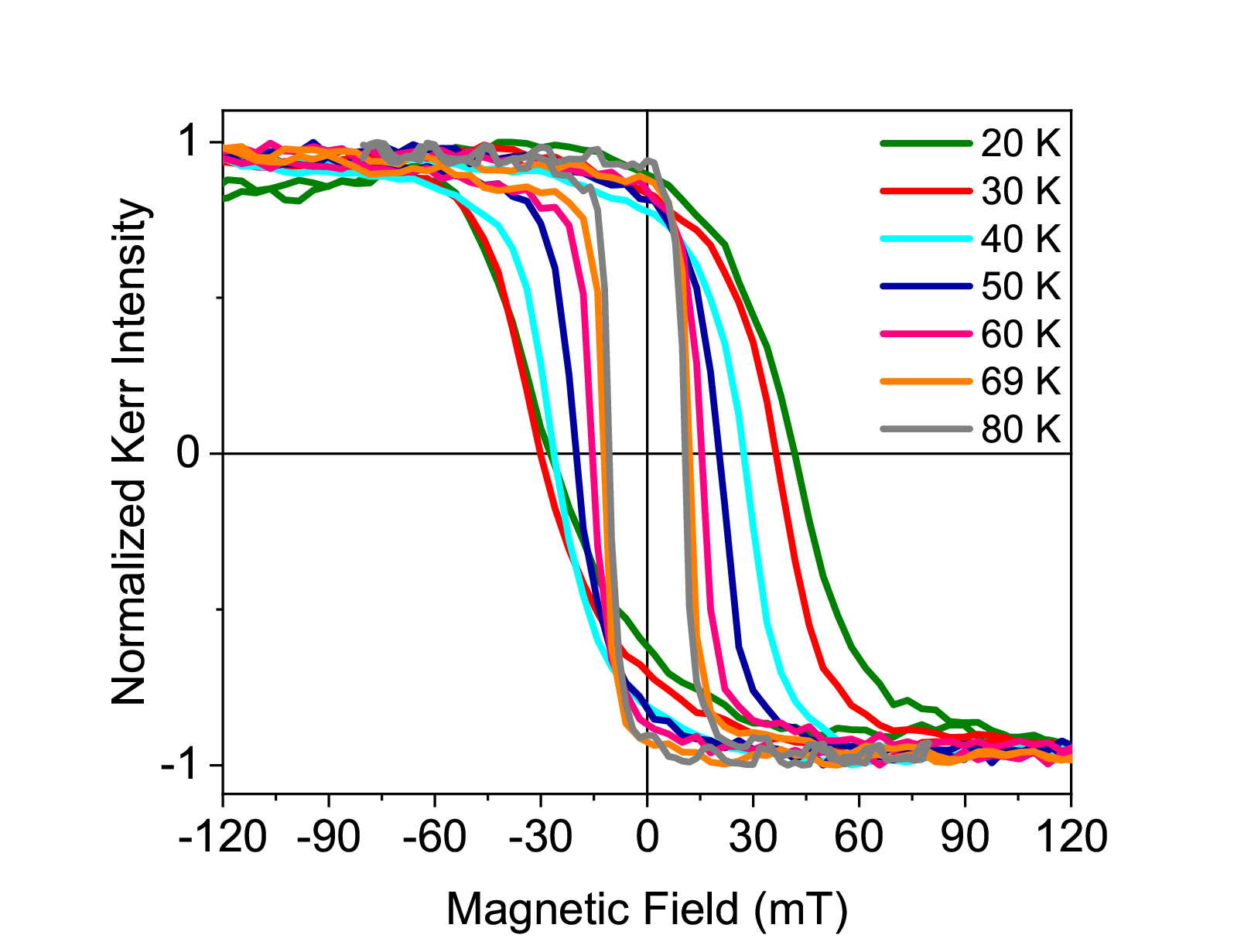}
\caption{\label{fig:fig3}Hystereses obtained using LMOKE at different temperatures on a 3\,nm EuO/4\,nmCo heterostructure.}
\end{figure}

Surprisingly different behavior was observed on the Co overlayer side of EuO/Co heterostructure. The temperature-dependent hystereses obtained are shown in Fig.~\ref{fig:fig3}. The hysteresis measurement was done simultaneously on both sides without changing the sample temperature to keep the variable parameters such as cooling/heating cycles to a minimum. At 30\,K and below, the hysteresis does not show a sudden jump to the magnetic saturation as was observed on the EuO side of the sample. On the Co side, the curves show a gradual saturation and possess asymmetric shape as well. Additionally, the Kerr contrast was significantly lower on the Co side and the loops were reversed. The hystereses also show a small exchange bias field ($\sim$\,5-6\,mT) at 10 and 20\,K. The \hc\,reduced similarly as the temperature is increased as observed for reference film and 3\,nm EuO. Such presence of exchange bias field in EuO films is generally attributed to the lack of stoichiometry \,\cite{offstoiEuO, JMMM_EB_EuO}. However, our in situ X-ray photoelectron spectroscopic investigations debar this as the EuO film is in stoichiometric phase.

Another possible explanation could be the presence of Co in the vicinity of EuO which leads to proximity coupling at the interface as observed recently in similar heterostructures \,\cite{HAXPES_MCD_EuO,dortmund2023}. This proximity-effect-induced coupling could shift the hysteresis and is found to be more pronounced below 30\,K in our case. Inferring from the observation of exchange bias, reversed and asymmetric loops without sudden saturation hinted that the nature of the coupling between EuO and Co is antiferromagnetic (AFM) type. Such AFM coupling between transition metals such as Fe, Co, etc., and rare earth layers has been already reported in the literature\,\cite{tang1998,okuno2020AFMbook,Goschew2016}. Such coupling is attributed to the hybridization between the \textit{3d} states of Co and the \textit{5d} states of Eu. This could also explain the observation of reversed loops in Fig.~\ref{fig:fig3}. As the Kerr rotation in EuO is significantly higher than that in Co ($\approx$\,0.3\dg\,and $\approx$\,2\dg\,for blue light), the coupling between them defines the dominant Kerr rotation. It seems that in our case, for EuO/Co system, Co dominates over EuO in the heterostructure unlike observed by M{\"o}nkeb{\"u}scher et. al.\,\cite{dortmund2023}. Here one thing that should be kept in mind is that the measurements had been performed using white light (400-700\,nm or 1.65-3.10\,eV).\\
For 40 and 50\,K, the hystereses approach to a more symmetric shape while \hc\,continues to decrease and similar behavior continues until 60\,K. The hysteresis curve becomes more symmetric afterward while \hc\,continues to reduce as temperature is increased which again gives hints that the coupling is either temperature-dependent or indirectly dependent on the temperature due to the temperature-dependent Kerr rotation of EuO. Domain images could not be observed on the Co capped EuO side, howsoever optical adjustments were made in the microscope. This could be due to very low Kerr contrast on the Co side of the sample and the AFM coupling diminished domains to a length or time scale which is beyond the scope of the measurement setup.

\subsection{\label{sec:HcvsT}Temperature-dependent \hc\,behavior}

\begin{figure}
\centering
\includegraphics[width=90mm]{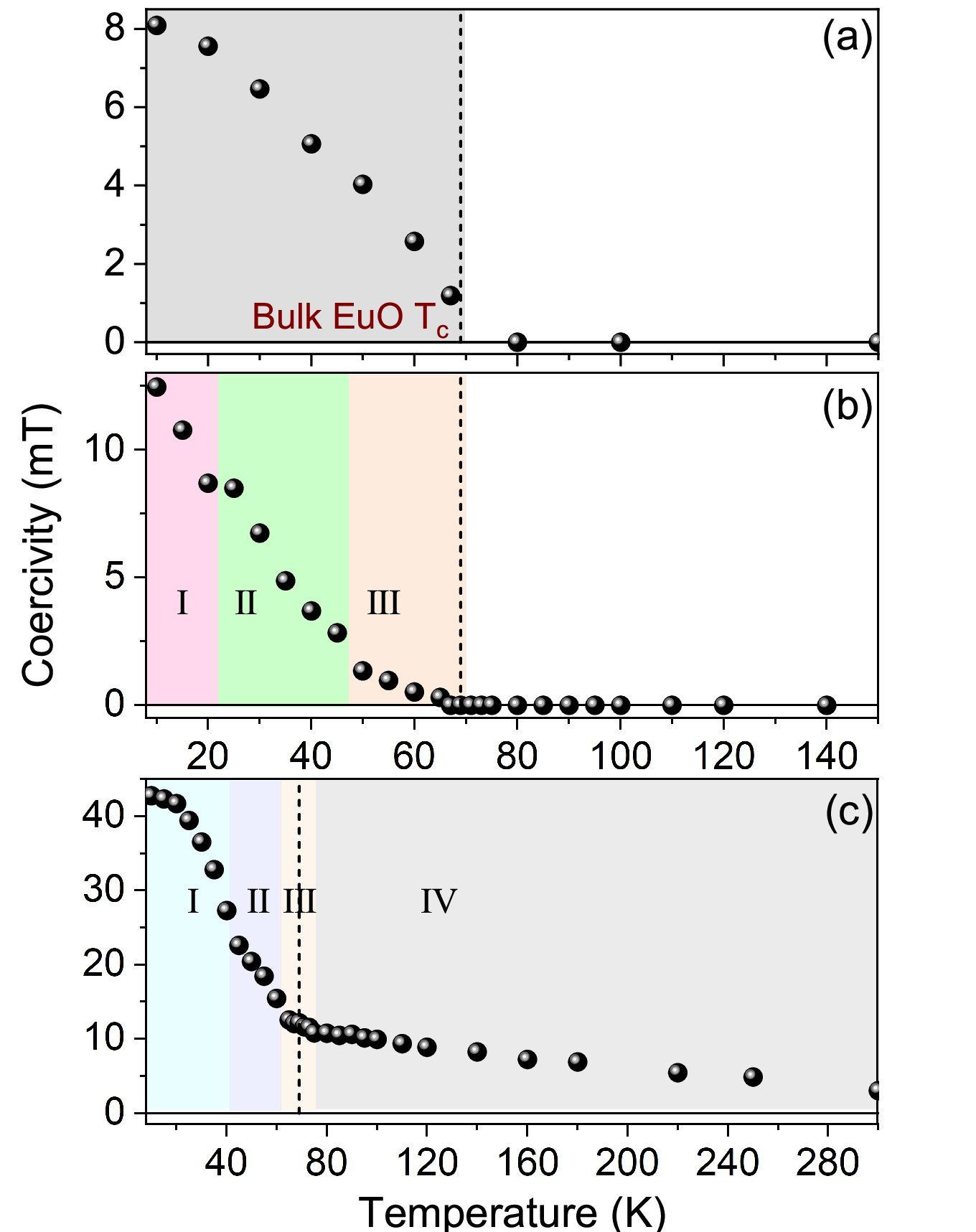}
\caption{\label{fig:fig4}The coercivity (\hc) dependence of temperature for (a) reference 25\,nm EuO film, (b) 3 nm EuO, and (c) Co(4\,nm)/EuO(3\,nm) heterostructure. The dashed vertical line is a guide to the eye for \Tc~of bulk EuO i.e. 69\,K. The error bars are less than the size of the symbols in the plot.}
\end{figure}

Fig.~\ref{fig:fig4} presents the \hc\,values obtained for the 25 nm EuO, 3 nm EuO, and 3\,nm EuO/4\,nm Co as the function of temperature. For the bulk-like reference EuO film, a monotonous decrease has been observed as temperature is increased. In the whole temperature range until the \Tc\,is reached, the behavior of decrement of \hc\,with temperature shows a typical shape as defined by Bloch's T$^{3/2}$~law for magnetization vs. temperature\,\cite{Brullion_FM_Heisenberg}. The trend of \hc\,variation can be understood by correlating the \hc\,with the Kerr rotation. The dependence of the Kerr signal for EuO films shows a similar behavior as reported by Swartz et.~al.\,\cite{APL_2010_EuO_GaAs}. However, in the Kerr microscope, we could not directly measure the Kerr rotation but only sense it indirectly with changes in the image intensity\,\cite{Schfer2007}. Since the \hc\,values here have been derived from LMOKE loops obtained by plotting Kerr intensity with the applied field, a similar dependence is to be expected.

In line with the differences observed in the hystereses and domain images of different thicknesses of EuO film, the \hc\,vs. temperature dependence is different for 3\,nm EuO. As can be seen in Fig.~\ref{fig:fig4}~(b), \hc\,decreases with the increase of the temperature in a different fashion. For 3\,nm EuO, the decrease of \hc\,can be understood by dividing the temperature range in  different regions. These three regions are I) low-temperature range (\begin{math}<\end{math}\,20\,K), II) intermediate temperatures (20\,K\,\begin{math}<\end{math}\,T\,\begin{math}<\end{math}45\,K), and III) nearby \Tc\,range. In region I, \hc\,decreases very sharply showing a Bloch's law-shaped dependence similar to reference film. In region II, the decrement of \hc\,shows a different trend as compared to the region I. Such dependence could stem from the formation of interfaces and their contribution to the overall \hc. As the film under observation is thin (3\,nm) as compared to the reference film (25\,nm), the contribution stemming from interfaces formed with the substrate and top MgO layer become more significant as compared to the reference film. From 50\,K until the point where Kerr signal could be obtained i.e. 65 K, the decrease in \hc\,is very small.

As we have already observed in Fig.~\ref{fig:fig3}, the deposition of the Co layer induced changes in the magnetic saturation behavior of the heterostructure. The temperature-dependent \hc\,behavior of EuO with Co along with region-wise variation is shown in Fig.~\ref{fig:fig4}~(c). Here, region I continues until 40\,K and the decrease of \hc\,with \Tc\,was less steep at first and then decreased more showing a dome-like Bloch law shape. After 40\,K, the variation of \hc\,showed a second dome-like behavior up to 60\,K. This could be attributed to one more interface coming into play i.e. between EuO and Co. In the critical region near \Tc\,of EuO, the shape of the hysteresis does not change anymore, giving the interpretation that in this temperature range, the signal mostly originates from the Co layer. The variations in \hc\,were very small and therefore, assigned as a separate region III. A clear distinction of the \Tc\,of EuO could not be seen as the Kerr signal from the Co layer start to dominate over EuO beyond its \Tc. Beyond region III, above the bulk \Tc\,of EuO, the hysteresis is originating only from Co layer and therefore the \hc\,variations could be seen until room temperature making region IV. Overall, region I to III constituted the effects of AFM interactions between EuO and Co and their interfaces, while region IV is the contribution of Co alone to the total \hc. In these measurements, the hystereses presented could not provide much information about the extent of proximity effect with temperature based on the changes in \hc\,as magneto-optical Kerr measurements lack element-specificity and practically limited Kerr signal resolution between the two layers.

\subsection{\label{sec:mumax}Micromagnetic Simulations}
\begin{figure*}[ht]
\centering 
\advance\leftskip-1cm
\includegraphics [width=140mm]{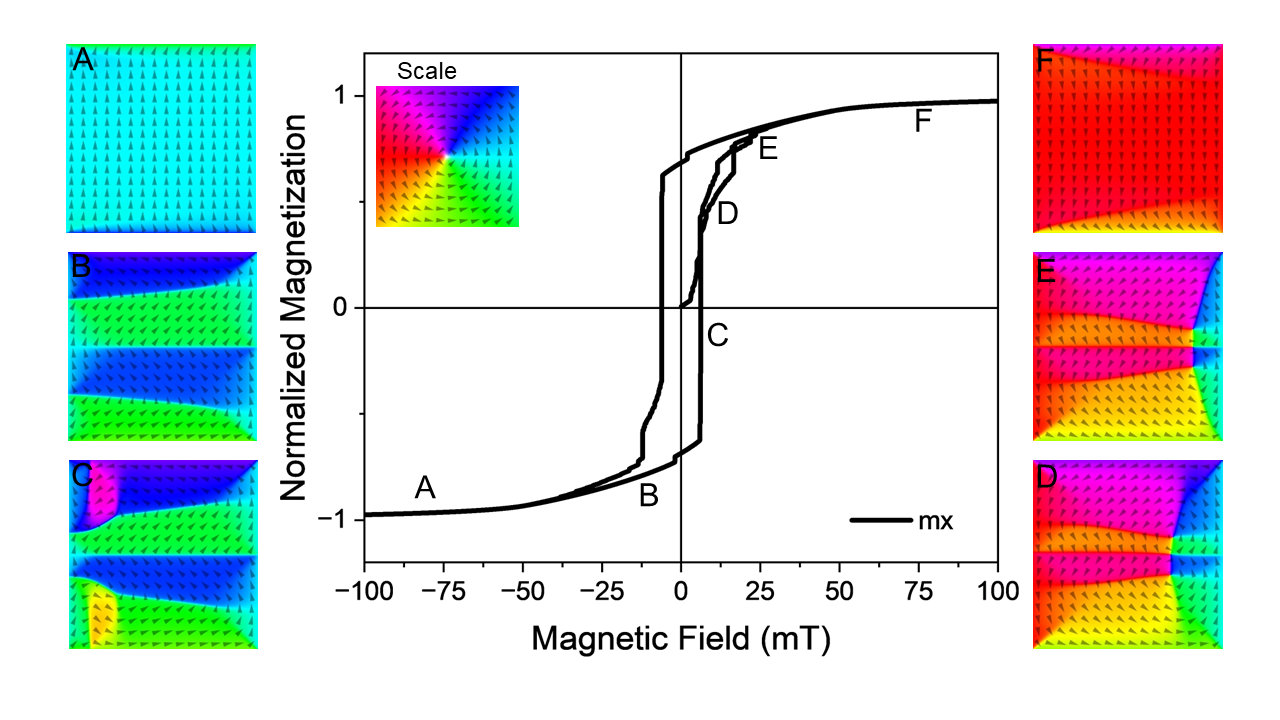}
\vspace{-7mm}
\caption{\label{fig:mumax1}Simulated magnetic hysteresis and different snapshots of magnetization in x-direction taken at various values of applied magnetic field for 3\,nm ultrathin EuO.  The corresponding color scale is shown as the insert in the hysteresis.}
\end{figure*}

A set of simulations was performed to check the possible transient domain configurations for the magnetic moments with the applied magnetic field in the x-direction. The normalized magnetization in the x, y, and z directions were recorded and plotted. It was found that with the given assumptions, magnetization is sensitive to x-direction only and a representative hysteresis is shown in Fig.~\ref{fig:mumax1}. The \hc~value for this loop is found to be 6\,mT, which is smaller than the experimental value for the 3\,nm EuO film. However, the snapshots of the magnetic spin configuration of the simulated loop showed similarities with the experimental results in section \ref{sec:EuOCo}. When the applied field was varied from -100\,mT, to +100\,mT, the snapshots taken at different field values (A--F) are also shown in Fig.~\ref{fig:mumax1}. At A, nearly all the spins are aligned with the applied magnetic field. When the the field was decreased, the moments attained a twin structure as shown in B. Such twin structures have been reported in literature~\cite{Prez2010}. When the crystallite size is small enough, such simple twin structure gets formed by alternating domains which touch each other through quasi-equidistant parallel plane 90\dg domain walls. Although we did not explicitly measure the crystallite size experimentally, however, it can be assumed to be small in a film of thickness 3\,nm.

The twin structures were observed in simulations only, not in the experimentally observed domain images. This can be understood by realizing the measurement sensitivity of the Kerr microscope. With the microscope, we only see two contrasts, which result from vector addition and cancellation on magnetic moments in pure LMOKE mode. In the simulations, we could observe them due to the distinct color resolution for each magnetization direction. Near the \hc, at point C, one could observe the regions emerging which are getting aligned in the direction of the applied field. This is in accordance with the experimental results, where small strip-like regions of aligned magnetic areas in the gray contrast could be observed. Further higher value of the the applied field at D results in an increase in area of the aligned magnetic moments along with twin structure formation as can be seen in E. From B to E, one can understand the magnetization reversal mechanism in EuO with the applied field. At F, nearly all the moments have aligned with the applied field and uniform magnetization could be observed. These simulated magnetic domain images corroborate the experimental observations  presented in the section~\ref{sec:EuOCo}. 
	
\section{\label{sec:conclusion}Conclusion}
In summary, we present the first magnetic domain imaging of bulk-like, ultrathin and proximity coupled films of EuO utilizing magneto-optical Kerr microscopy. The focus was on the evolution of magnetic domains below and up to \Tc. Our observations revealed that the Kerr contrast and domain size in EuO are strongly dependent on the film thickness. Specifically, we observed well-defined, high-contrast magnetic domains in a 25\,nm thick EuO film down to the ultrathin film limit (3\,nm). As the thickness of the EuO film decreased, the size of magnetic domains was found to diminish. Simultaneous domain imaging and hysteresis measurements enabled us to analyze the temperature-dependent behavior of \hc. We discovered that variations in \hc~with temperature are influenced by film thickness as well, with interfacial contributions becoming more prominent in thinner EuO films. The presence of a Co overlayer on EuO in the heterostructure resulted in asymmetric hysteresis, characterized by relatively high \hc. Additionally, the antiferromagnetic proximity coupling between EuO and Co significantly influenced both \hc~and magnetic domains. Micromagnetic simulations conducted for a 3\,nm EuO film exhibited consistency with experimental observations obtained through pure longitudinal MOKE measurements. Our findings underscore the potential of designing and conducting fundamental studies on EuO and 3d transition metal-based heterostructures and could lead to the realization of sustained magnetic ordering in EuO beyond its critical temperature.

\section{\label{sec:ack}Acknowledgement}
The authors acknowledge SFB\,1432 with Project-ID425217212, University of Konstanz for funding (Project B03), and Prof. Dr. R. Sch{\"a}fer (IFW Dresden) for insightful discussions. Seema thank Ricardo Pérez Twerdy for the help in simulations and University of Konstanz for RiSC-program project funding. 

\section{\label{sec:ack}Data availability}
The datasets used and/or analyzed during the current
study available from the corresponding author on reasonable request.

\bibliography{EuO_ref}

\pagebreak
\widetext
\begin{center}
\subsection{{Supplementary information:} Magnetic domains in ultrathin, bulk-like and proximity-coupled Europium Oxide}
	\maketitle	
\end{center}
	\subsubsection{\label{sec:exps}Experimental}
	The mechanism to observe domains with a Kerr microscope is based on the method of taking background image in a magnetically saturated state. Here, the sample is first saturated in a certain value of applied field and a background image is taken. This provides fully a saturated single domain as a gray image. The negative/positive sign of the applied field symbolizes the direction of the field. Then the background image is subtracted from the live image which is captured at each value of applied field of the hysteresis. The intensity of this subtracted image at each field value is used to plot the hysteresis curve. Therefore, near the coercive field when the domains switch their direction, their reversal state can be captured as a subtracted image of live and single domain background image.
	
	The pure LMOKE approach was needed to obtain enhanced in-plane signal, symmetric, and comparable loops on both sides of the EuO/Co heterostructure. The microscope provides an option for fine-tuning in-plane Kerr sensitivity by making light incident from opposite sides and subtracting the corresponding live images. This process helps to cancel out any out-of-plane or transverse components and enhance only in-plane sensitivity by superposition as discussed in detail by Soldatov et al. and is useful for the samples with weak in-plane magnetization due to some out-of-plane contributions\,\cite{Soldatov_advancedKerr}.
	
	The Kerr intensity of a chosen sample area as a function of the magnetic field is used to plot the hysteresis loops. After being normalized to 1, the recorded hystereses in the objective lens were adjusted for drift and Faraday effects using the software\,\cite{Schfer2007}. A Cryovac sample stage that can be cooled to 4\,K was used since EuO is ferromagnetic below 69\,K only. Helium flow type cryostat was used to cool the thin film sample which was placed on a cold finger in an isovacuum of 10$^{-6}$\,Torr. The upper section of this sample stage has a glass window that permits incident and reflected light. An objective lens specifically designed to rectify the impact of a glass window on reflected light was used, boasting the image magnification to 60x. Temperature stabilization was attained using a TIC500 power supply from Cryovac where temperature stability up to 0.001\,K can be achieved.
	
	For the EuO/Co heterostructure, Co has been deposited on half of the EuO sample. Therefore, simultaneous measurements were carried out on the EuO side and the EuO/Co side at each temperature and in a single cooling cycle. The sample was translated left and right for these measurements in the pure LMOKE mode. The boundary between two regions (EuO with Co layer and without Co layer) was optically visible in the microscopic image and utmost care was taken in the measurements for area selection from which the Kerr signal was recorded. This was performed by keeping an eye on some surface defects and using them as location markers.
	
	To realize a micromagnetic model of EuO, we need to define number of cells in the x, y, and z directions. In the present case, we assumed the simulation grid size to be 256$\times$256$\times$6, while the size of the cell was taken to be 0.5\,nm in x, y, and z direction, respectively. In this way, the thickness of the model film is 3\,nm, resembling the thin EuO sample. To allow the exchange coupling between the magnetic moments contained in the neighboring cells, the cell edge length has to be smaller than the exchange length (\textit{l}$_{exch}$), which can be calculated as $l_{exch}=\sqrt{\frac{A_{exch}}{K_C}}$. In this way, the magnetization going from one cell to another does not undergo sharp changes. The evolution of the magnetization is evaluated as its time derivative of torque in Mumax$^3$, which solves the Landau-Lifshitz-Gilbert equation\,\cite{Leliaert2018}. Periodic boundary conditions were not taken in to account in the presented simulated loop. The simulations were performed at 30\,K using simple script written in Golang and executing them at the GPU nodes at HPC cluster SCCKN
	of the University of Konstanz.
	
	\clearpage
	
	\subsubsection{Results}
	\begin{figure} [ht]
		\centering
		\includegraphics [width=105mm]{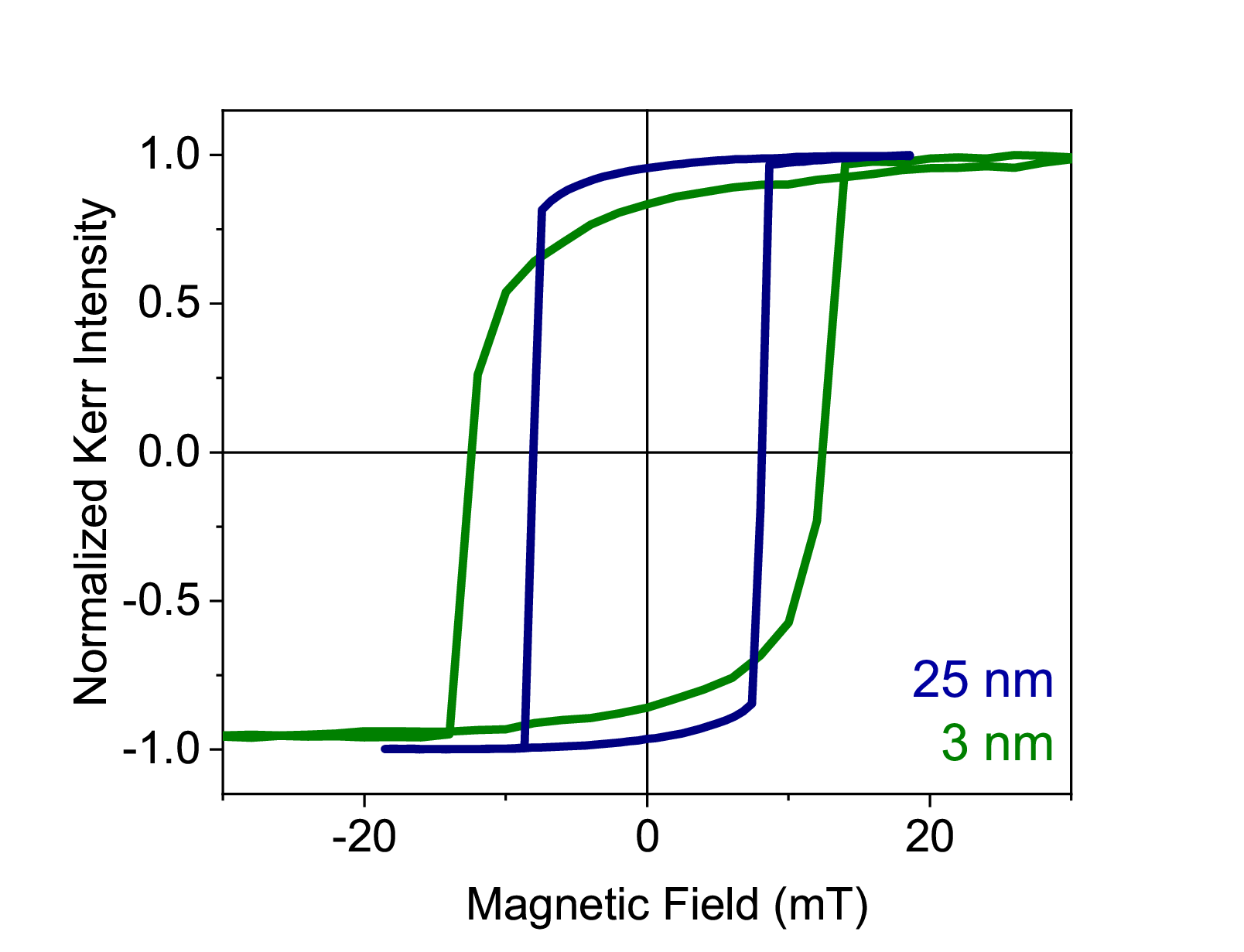}
		\caption{\label{fig:fig1S}Hysteresis obtained for 25\,nm and 3\,nm EuO film at 10\,K using LMOKE and pure LMOKE, respectively.}
	\end{figure}
	
	\begin{figure*}[ht]
		\centering
		\begin{tikzpicture}
			\node[inner sep=0pt] (image) at (0,0) {
				\includemedia[
				width=0.95\linewidth,
				height=0.75\linewidth,
				activate=pageopen,
				addresource=domainvideo25nm.mp4,
				flashvars={source=domainvideo25nm.mp4
					&autoPlay=true
					&loop=true
				}
				]{}{VPlayer.swf}};
			\draw[line width=1pt,black] (image.south west) rectangle (image.north east);
		\end{tikzpicture}
		\captionof{figure}{The magnetic domain evolution in form of stripes in the reference 25\,nm EuO film recorded separately by varying the applied magnetic field and varying it manually very slowly. The measurements were performed in LMOKE geometry.} 
		\label{fig:vid25}
	\end{figure*}
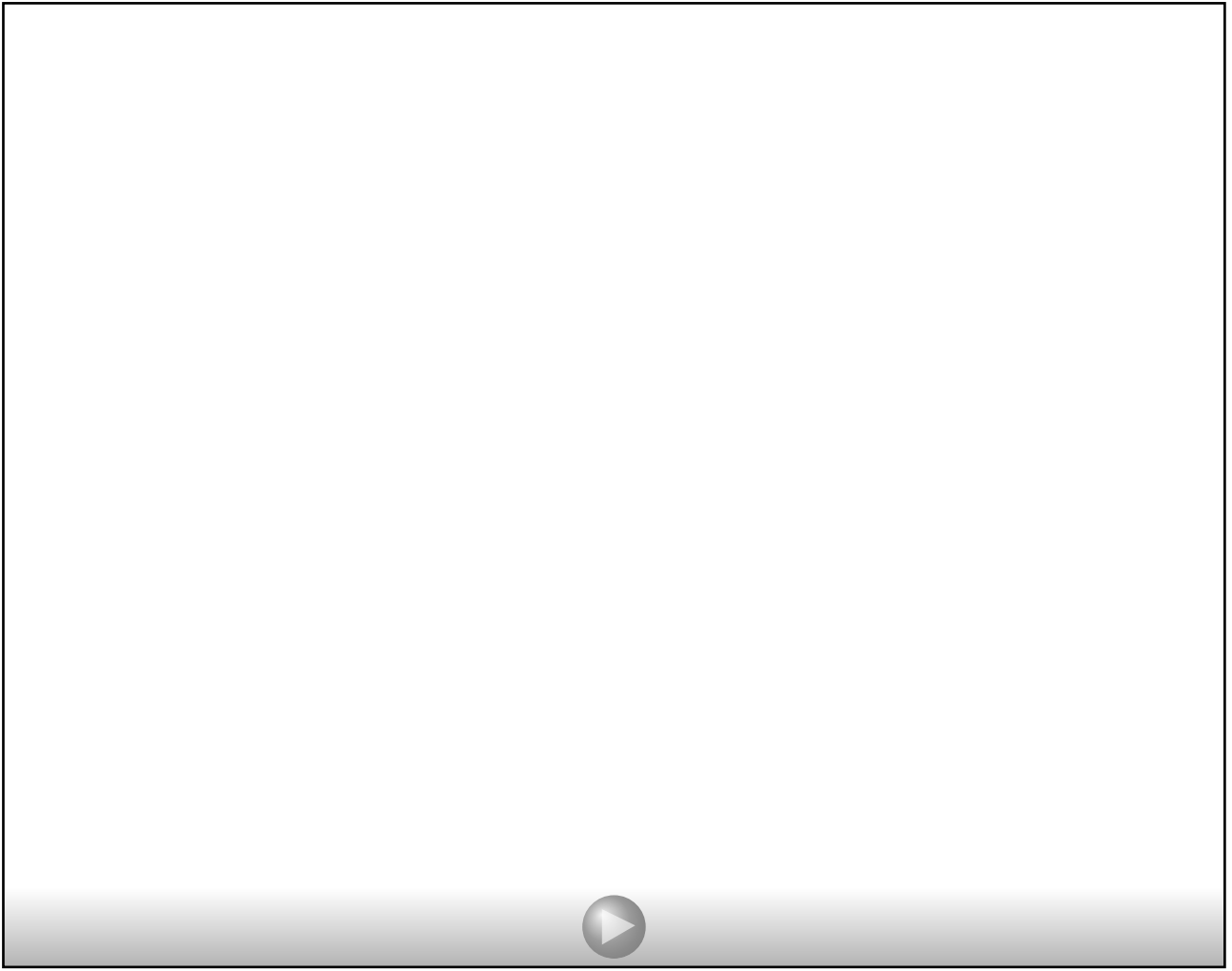
	
	\begin{figure*}[ht]
		\centering
		\begin{tikzpicture}
			\node[inner sep=0pt] (image) at (0,0) {
				\includemedia[
				width=0.95\linewidth,
				height=0.75\linewidth,
				activate=pageopen,
				addresource=domainvideo3nm.mp4,
				flashvars={source=domainvideo3nm.mp4
					&autoPlay=true
					&loop=true
				}
				]{}{VPlayer.swf}};
			\draw[line width=1pt,black] (image.south west) rectangle (image.north east);
		\end{tikzpicture}
		\captionof{figure}{The magnetic domain evolution in form of small stripes in ultrathin 3\,nm EuO film recorded separately by varying the applied magnetic field and varying it manually very slowly. The measurements were performed in the pure LMOKE geometry as explained in the experimental section~\ref{sec:exps}.} 
		\label{fig:vid3}
	\end{figure*}
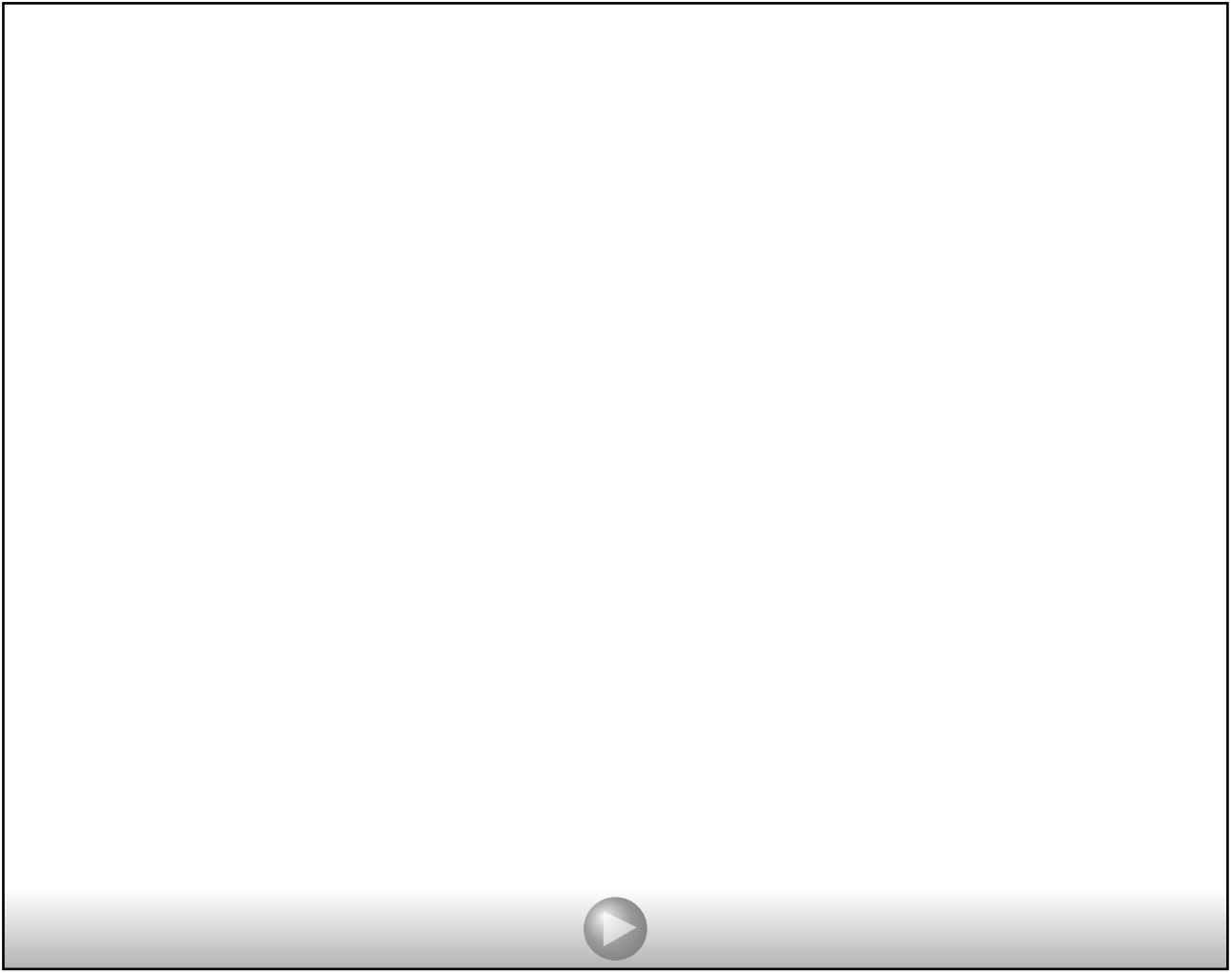

\end{document}